# Treemaps with Bounded Aspect Ratio[☆]

Mark de Berg[a], Bettina Speckmann[a], Vincent van der Weele[b]

[a]*Department of Mathematics and Computer Science, TU Eindhoven, The Netherlands*
[b]*Max-Planck-Institut für Informatik, Saarbrücken, Germany*

**Abstract**

Treemaps are a popular technique to visualize hierarchical data. The input is a weighted tree $\mathcal{T}$ where the weight of each node is the sum of the weights of its children. A treemap for $\mathcal{T}$ is a hierarchical partition of a rectangle into simply connected regions, usually rectangles. Each region represents a node of $\mathcal{T}$ and its area is proportional to the weight of the corresponding node. An important quality criterion for treemaps is the aspect ratio of its regions. One cannot bound the aspect ratio if the regions are restricted to be rectangles. In contrast, *polygonal partitions*, that use convex polygons, have bounded aspect ratio. We are the first to obtain convex partitions with optimal aspect ratio $O(\text{depth}(\mathcal{T}))$. However, $\text{depth}(\mathcal{T})$ still depends on the input tree. Hence we introduce a new type of treemaps, namely *orthoconvex treemaps*, where regions representing leaves are rectangles, L-, and S-shapes, and regions representing internal nodes are orthoconvex polygons. We prove that any input tree, irrespective of the weights of the nodes and the depth of the tree, admits an orthoconvex treemap of constant aspect ratio. We also obtain several specialized results for single-level treemaps, that is, treemaps where the input tree has depth 1.

## 1. Introduction

Treemaps are a very popular technique to visualize hierarchical data [14]. The input is a tree $\mathcal{T}$ where every leaf is associated with a weight and where the weight of an internal node is the sum of the weights of its children. A treemap for $\mathcal{T}$ is a hierarchical partition of a simple polygon, usually a rectangle, into simply connected regions, often rectangles as well. Each such region represents a node of $\mathcal{T}$ and the area of each region is proportional to the weight of the corresponding node. To visualize the hierarchical structure the region associated with a node must contain the regions associated with its children. Shneiderman [15] and his colleagues were the first to present an algorithm for the automatic creation of rectangular treemaps. Treemaps have since been used to

[☆]Bettina Speckmann was supported by the Netherlands' Organisation for Scientific Research (NWO) under project no. 639.022.707.

visualize hierarchical data from a variety of application areas, for example, stock market portfolios [10], tennis competitions trees [9], large photo collections [3], and business data [16].

One of the most important quality criteria for treemaps is the aspect ratio of its regions; users find it difficult to compare regions with extreme aspect ratios [11]. Hence several approaches [3, 4] try to "squarify" the regions of a rectangular treemap. However, one cannot bound the aspect ratio if the regions are restricted to be rectangles. (Consider a tree con-sisting of two leaves and a root and let the weight of one leaf tend to zero.) As a consequence, several types of treemaps using region shapes other than rectangles have been proposed. Balzer and Deussen [1, 2] use centroidal Voronoi tessellations. Their algorithm is iterative and can give no guarantees on the aspect ratio of the regions (nor on their exact size). Wattenberg [17] developed treemaps whose regions follow a space filling curve on a grid, so called Jigsaw maps. Jigsaw maps assume that the leaves have integer weights which add up to a square number. The regions of the maps are rectilinear, but highly non-(ortho)convex. However, they do have aspect ratio 4. Onak and Sidiropoulos [13] introduced *polygonal partitions*, which use convex polygons. They proved an aspect ratio of $O((\text{depth}(\mathcal{T}) \cdot \log n)^{17})$ for a tree $\mathcal{T}$ with $n$ leaves. In cooperation with De Berg, this bound has since been improved to $O(\text{depth}(\mathcal{T}) + \log n)$ [5]. The latter paper also gives a lower bound of $\Omega(\text{depth}(\mathcal{T}))$.

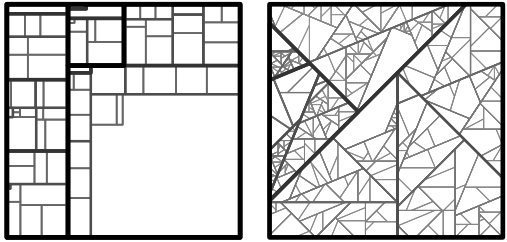

Figure 1: Treemaps constructed by our drawing algorithms: orthoconvex and convex.

This leaves two open questions. First, is the $O(\log n)$ term in the upper bound on convex treemaps necessary, or can we guarantee a $O(\text{depth}(\mathcal{T}))$ aspect ratio? Second, is there a type of treemap that has constant aspect ratio for any input tree, irrespective of its depth and of the number of nodes and their weights?

**Results and organization.** We answer the two questions above affirmatively. First of all, in Section 3, we show how to construct convex partitions with optimal aspect ratio $O(\text{depth}(\mathcal{T}))$. Second, we introduce *orthoconvex treemaps*, where the regions representing internal nodes of the input tree are ortho-convex polygons, while the regions representing leaves are rectangles, L-shapes, and S-shapes (see Fig. 2). Thus our orthoconvex treemaps retain some of the schematized flavor of rectangular treemaps. In Section 4 we prove that any input tree admits an orthoconvex treemap of constant aspect ratio. Fig. 1 shows two treemaps constructed by our drawing algorithms. The hierarchy is emphasized by line thickness and color: thicker, darker lines delimit nodes higher in the hierarchy.

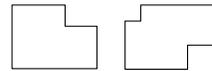

Figure 2: An L- and an S-shape (two reflex corners).

We also consider the special case that depth($\mathcal{T}$) = 1, that is, single-level treemaps. For rectangular treemaps, minimizing the aspect ratio is NP-hard for single-level treemaps. This result was claimed without proof in [4], but was in fact only a conjecture of the authors[1]. For completeness, we give a formal proof in Section 5.1. In Section 5.2 we describe a drawing procedure that uses L-shapes in addition to rectangles. The resulting treemap has aspect ratio at most $2 + 2\sqrt{3}/3 \approx 3.15$. We also give a construction which forces a maximal aspect ratio of $\approx 3.13$ for such treemaps in Section 5.3.

Instead of using L-shapes we can also draw a single-level treemap using four different convex octilinear shapes. The resulting treemap has aspect ratio at most $\approx 4.79$. However, since the principle of the construction is very similar to that in Section 3 and the aspect ratio bound is far from tight, we do not include any further details on this construction.

## 2. Preliminaries

Our input is a rooted tree $\mathcal{T}$. Following [5] we say that $\mathcal{T}$ is *properly weighted* if each node $\nu$ of $\mathcal{T}$ has a positive weight weight($\nu$) that equals the sum of the weights of the children of $\nu$. We assume that the weights are normalized, that is, weight(root($\mathcal{T}$)) = 1. A treemap for $\mathcal{T}$ associates a region $R(\nu)$ with each node $\nu \in \mathcal{T}$ such that (i) $R(\text{root}(\mathcal{T}))$ is the unit square, (ii) for every node we have area($R(\nu)$) = weight($\nu$), and (iii) for any node $\nu$, the regions associated with the children of $\nu$ form a partition of $R(\nu)$.

The aspect ratio of a treemap is the maximum aspect ratio of any of its regions. To simplify our calculations, we use a different definition of aspect ratio for orthoconvex and for convex regions. Let $R$ be a region and $\sigma(R)$ its smallest enclosing axis-aligned square. Furthermore, let area($R$) be its area and diam($R$) its 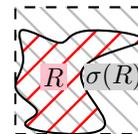
diameter. For orthoconvex regions, we define the aspect ratio of $R$ as $\text{asp}(R) := \text{area}(\sigma(R))/\text{area}(R)$. For convex regions, we define it in accordance with [5] as $\text{asp}(R) := \text{diam}(R)^2/\text{area}(R)$. Note that $\text{area}(\sigma(R)) \leq \text{diam}(R)^2 \leq 2 \cdot \text{area}(\sigma(R))$, so the aspect ratios obtained by the two definitions differ by at most a factor 2.

The following two lemmas deal with partitioning the children of a node according to weight and with partitioning a rectangular region. We denote the set of children of a node $\nu$ by children($\nu$).

**Lemma 1.** *Suppose all children of node $\nu$ have weight at most $t \cdot \text{weight}(\nu)$, for some $3/10 \leq t \leq 2/3$. Then we can partition* children($\nu$) *into two subsets $H_1$ and $H_2$, such that*

$$\text{weight}(H_2) \leq \text{weight}(H_1) \leq \begin{cases} 2t \cdot \text{weight}(\nu) & \text{if} \quad 3/10 \leq t < 1/3; \\ 2/3 \cdot \text{weight}(\nu) & \text{if} \quad 1/3 \leq t \leq 2/3. \end{cases}$$

---

[1]Personal communication with J.J. van Wijk and K. Huizing, Nov 2010.

PROOF. We partition the children using the *longest processing time* (LPT) algorithm [8]. The algorithm works as follows: given a list of weights $[w_1, w_2, \ldots w_n]$ in non-increasing order, and two bins $H_1, H_2$, repetitively place the highest weight left in the bin with least total weight. Consider the last weight $w^*$ put in bin $H_1$. By LPT, $H_1$ was the emptiest bin before $w^*$ was put in, hence $\text{weight}(H_1) - w^* \leq \text{weight}(H_2)$ and thus $\text{weight}(H_1) \leq (\text{weight}(\nu) + w^*)/2$.

If $1/3 \leq t \leq 2/3$, we distinguish two cases. If $1/2 \leq w_1/\text{weight}(\nu) \leq 2/3$, then it is clear that $\text{weight}(H_1) = w_1 \leq 2/3 \cdot \text{weight}(\nu)$. If $1/3 \leq w_1/\text{weight}(\nu) < 1/2$, then $w_3$ is the first weight that can cause $\text{weight}(H_1) \geq 1/2 \cdot \text{weight}(\nu)$. The weights are in non-increasing order, hence $w_i \leq \text{weight}(\nu)/i$. Therefore, $w^* \leq \text{weight}(\nu)/3$ and $\text{weight}(H_1) \leq 2/3 \cdot \text{weight}(\nu)$.

If $3/10 \leq t < 1/3$, we distinguish three cases. If $w_2 + w_3 \geq \text{weight}(\nu)/2$, then $H_1 = \{w_2, w_3\}$. It is not hard to see that $H_1$ has maximal weight if $w_3 = w_2 = w_1 = t$, hence $\text{weight}(H_1) \leq 2t \cdot \text{weight}(\nu)$. If $w_2 + w_3 < \text{weight}(\nu)/2$ but $w_1 + w_4 \geq \text{weight}(\nu)/2$, then $\text{weight}(H_1) < \text{weight}(\nu)/3 + \text{weight}(\nu)/4 = 7/12 \cdot \text{weight}(\nu) < 2t \cdot \text{weight}(\nu)$. If $w_2 + w_3 < \text{weight}(\nu)/2$ and $w_1 + w_4 < \text{weight}(\nu)/2$, then $w^* \leq \text{weight}(\nu)/5$, hence $\text{weight}(H_1) \leq 3/5 \cdot \text{weight}(\nu) \leq 2t \cdot \text{weight}(\nu)$. □

**Lemma 2.** *Let $R$ be a rectangle and $w_1 \geq w_2$ be weights such that $w_1 + w_2 = \text{area}(R)$. Then we can partition $R$ into two subrectangles $R_1, R_2$, such that $\text{asp}(R_i) \leq \max(\text{asp}(R), \text{area}(R)/w_i)$, for $i \in \{1, 2\}$.*

PROOF. $R_i$ has aspect ratio $\max(w_i \cdot \text{asp}(R)/\text{area}(R), \text{area}(R)/(w_i \cdot \text{asp}(R)))$, for $i \in \{1, 2\}$. The claim follows from $w_i/\text{area}(R) \leq 1$ and $1/\text{asp}(R) \leq 1$. □

## 3. Convex Treemaps

We describe a recursive algorithm for computing a convex treemap (polygonal partition) of aspect ratio $O(\text{depth}(\mathcal{T}))$ for a properly weighted tree $\mathcal{T}$. Our algorithm has two phases. We first convert $\mathcal{T}$ into a binary tree $\mathcal{T}^*$ and then construct a partition for $\mathcal{T}^*$. Roughly speaking, at every step our algorithm finds a line to split a given convex polygon with "good" aspect ratio according to the weights of the two children of the current node. Both sub-polygons have good aspect ratio again. We cut with axis-aligned lines whenever we can. In fact we can ensure that we introduce new cutting directions only when encountering a new level of the original input tree instead of on every level of the binary tree. This is the key to obtaining $O(\text{depth}(\mathcal{T}))$ aspect ratio, rather than $O(\log n + \text{depth}(\mathcal{T}))$.

**Converting to a binary tree.** We recursively convert $\mathcal{T}$ into a strictly binary tree $\mathcal{T}^*$, replacing each node with $k > 2$ children in $\mathcal{T}$ by a binary subtree with $k - 1$ new internal nodes. Since a binary tree with $k - 1$ internal nodes can accommodate exactly $k$ leaves, the children of the original internal node can all become a child of one of the new internal nodes. During this process we assign a label $d(\nu)$ to each node $\nu$, which corresponds to the depth of $\nu$ in $\mathcal{T}$. In a generic step, we treat a node $\nu$ with label $d(\nu)$, and our task is to convert the subtree rooted at $\nu$. Initially $\nu = \text{root}(\mathcal{T})$ with $d(\text{root}(\mathcal{T})) = 0$. If $\nu$ is a leaf there is nothing to do. If $\nu$ has two children we recurse on these children and assign them label $d(\nu) + 1$. Otherwise $\nu$ has $k$ children, children$(\nu) = \{\nu_1, \ldots, \nu_k\}$, for some $k > 2$. We distinguish two cases, depending on their weight.

If there is a "heavy" child, say $\nu_1$, such that weight$(\nu_1) \geq$ weight$(\nu)/2$, then we proceed as follows. We turn $\nu$ into a binary node whose children are $\nu_1$ and a new node $\mu_1$; the children of $\mu_1$ are $\nu_2, \ldots, \nu_k$. We recurse on $\nu_1$ and on $\mu_1$, with $d(\nu_1) = d(\nu) + 1$ and $d(\mu_1) = d(\nu)$. Otherwise all children have weight less than weight$(\nu)/2$, and hence there is a partition of children$(\nu)$ into two subsets $S_1$ and $S_2$ such that weight$(S_i) \leq 2/3 \cdot$ weight$(\nu)$ for $i \in \{1, 2\}$. We turn $\nu$ into a binary node with children $\mu_1$ and $\mu_2$, with children from $S_1$ and $S_2$, respectively, and we recurse on $\mu_1$ and $\mu_2$ with $d(\mu_1) = d(\mu_2) = d(\nu)$.

**Drawing a binary tree.** Generalizing $\phi$-separated polygons [5], we define a $(k, \phi)$-*polygon*, with $k \geq 1$, to be a convex polygon $P$ such that
  (i) $P$ does not have parallel edges, except possibly two horizontal edges and two vertical edges. Moreover, each non-axis-parallel edge $e$ makes an angle of at least $\phi$ with any other edge and also with the $x$-axis and the $y$-axis.
 (ii) If $P$ has two horizontal edges, then width$(P)/$ height$(P) \leq k$.
(iii) If $P$ has two vertical edges, then height$(P)/$ width$(P) \leq k$.

A $\phi$-separated polygon can only have axis-parallel edges if these are parts of its axis-aligned bounding square [5]. Therefore, a $(k, \phi)$-polygon $P$ is a $\phi$-separated polygon, if it respects the following:
  (a) if $P$ has two horizontal edges, then height$(P) \geq$ width$(P)$;
  (b) if $P$ has two vertical edges, then width$(P) \geq$ height$(P)$.

Note that a $(k, \phi)$-polygon $P$ is $\phi$-separated if its bounding box is square.

**Lemma 3.** *Any $(k, \phi)$-polygon has aspect ratio $O(\max(k, 1/\phi))$.*

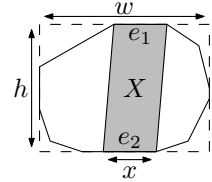

PROOF. Consider a $(k, \phi)$-polygon $P$. For brevity, we write $w = $ width$(P)$ and $h = $ height$(P)$. Assume, without loss of generality, that $w \geq h$. Let $e_1$ and $e_2$ be the horizontal edges (possibly of length 0) and let $x = \min(|e_1|, |e_2|)$. Let $X$ be the shaded parallellogram of width $x$. We distinguish two cases.

**Case 1:** $x > w/2$. $P$ has two horizontal edges, so $h \geq w/k$. Clearly, area$(P) \geq$ area$(X)$ which is $xh > w^2/(2k)$. The diameter of $P$ is at most the diameter of the enclosing rectangle, hence diam$(P)^2 \leq w^2 + h^2 \leq 2w^2$. Combined:

$$\mathrm{asp}(P) = \frac{\mathrm{diam}(P)^2}{\mathrm{area}(P)} \leq \frac{2w^2}{w^2/(2k)} = 4k = O(k)\,.$$

**Case 2:** $x \leq w/2$. We obtain polygon $P'$ from $P$ by reducing the length of $e_1$ and $e_2$ by $\min(x, w - h)$. Clearly, area$(P') \leq$ area$(P)$. Observe that $P'$ is a $\phi$-separated polygon since either it has at most 1 horizontal edge (a) and $w - x \geq h$ (b), or the bounding box of $P'$ is square. Therefore, asp$(P') = O(1/\phi)$ [5]. Using $\text{diam}(P) \leq w\sqrt{2}$ and $\text{diam}(P') \geq w - \min(x, w - h) \geq w - x \geq w/2$, we calculate

$$\text{asp}(P) = \frac{\text{diam}(P)^2}{\text{area}(P)} \leq \frac{2w^2}{\text{area}(P)} \leq 8 \cdot \frac{\text{diam}(P')^2}{\text{area}(P')} = 8 \cdot \text{asp}(P') = O(1/\phi). \quad \square$$

We construct the partition for $\mathcal{T}^*$ in a top-down manner. Suppose we arrive at a node $\nu$ in $\mathcal{T}^*$, with associated region $R(\nu)$; initially $\nu = \text{root}(\mathcal{T}^*)$ and $R(\nu)$ is the unit square. We write $n(\nu)$ for the number of non-axis-parallel edges in $R(\nu)$. We maintain the following invariants:

**(Inv-1)** $n(\nu) \leq d(\nu) + 4$;
**(Inv-2)** $R(\nu)$ is a $(k, \phi(\nu))$-separated polygon for $k = 4$ and $\phi(\nu) = \pi/(2(d(\nu) + 6))$.

Note that the invariant is satisfied for $\nu = \text{root}(\mathcal{T}^*)$. Now consider a node $\nu$ that is not the root of $\mathcal{T}^*$. If $\nu$ is a leaf, there is nothing to do. Otherwise, let $\nu_1$ and $\nu_2$ be the two children of $\nu$. Assume without loss of generality that $\text{weight}(\nu_1) \geq \text{weight}(\nu_2)$. We distinguish two cases.

**Case 1:** $d(\nu_1) = d(\nu) + 1$. We consider the lines parallel to the edges of $R(\nu)$ through the origin. Moreover, we add the x- and y-axis. Since $R(\nu)$ has at most $d(\nu) + 4$ non-axis-parallel edges, we have at most $d(\nu) + 6$ lines in total. Hence, the biggest gap between two subsequent lines is at least $\pi/(d(\nu)+6)$. Therefore, there is a line $\ell$ that makes an angle of at least $\pi/(2(d(\nu) + 6))$ with each of the edges of $R(\nu)$ and with the x- and the y-axis. Imagine placing the line $\ell$ such that it splits $R(\nu)$ into two halves of equal area, and define $R'$ to be the half with the smallest number of non-axis-parallel edges. Now partition $R(\nu)$ into subpolygons $R(\nu_1)$ and $R(\nu_2)$ of the appropriate area with a cut $c$ that is parallel to $\ell$ such that $R(\nu_2) \subset R'$. (Thus $c$ lies inside $R'$.) We claim that both $R(\nu_1)$ and $R(\nu_2)$ satisfy the invariant.

Clearly $R(\nu_1)$ uses at most one edge more than $R(\nu)$. Since $d(\nu_1) = d(\nu)+1$, this implies that (Inv-1) is satisfied for $R(\nu_1)$. Now consider the number of non-axis-parallel edges of $R(\nu_2)$. This is no more than the number of non-axis-parallel edges of $R'$. At most two non-axis-parallel edges are on both sides of $\ell$, hence this number is bounded by

$$n(\nu_2) \leq \left\lfloor \frac{n(\nu) + 2}{2} \right\rfloor + 1 \leq \left\lfloor \frac{d(\nu) + 6}{2} \right\rfloor + 1 = \left\lfloor \frac{d(\nu)}{2} \right\rfloor + 4 \leq d(\nu) + 4 \leq d(\nu_2) + 4.$$

Given the choice of $\ell$, and because $d(\nu_i) \geq d(\nu)$ and $R(\nu)$ satisfies (Inv-2), we know that the minimum angle between any two non-parallel edges of $R(\nu_i)$ ($i \in \{1, 2\}$) is at least $\pi/(2(d(\nu_i) + 6))$. To show that $R(\nu_1)$ and $R(\nu_2)$ satisfy (Inv-2), it thus suffices to prove the following lemma.

**Lemma 4.** *If $R(\nu_i)$ has two horizontal edges, then $\text{width}(R(\nu_i))/\text{height}(R(\nu_i)) \leq k$ and if $R(\nu_i)$ has two vertical edges, then $\text{height}(R(\nu_i))/\text{width}(R(\nu_i)) \leq k$, for $i \in \{1, 2\}$.*

PROOF. We prove only the first claim, the second proof is similar. Assume $R(\nu_i)$ has two horizontal edges $e_1$ and $e_2$. Cut $c$ is neither horizontal nor vertical, hence $e_1$ and $e_2$ show also up as (parts of) edges in $R(\nu)$. Hence, $\text{height}(R(\nu_i)) = \text{height}(R(\nu))$. Since $R(\nu_i) \subset R(\nu)$, we know $\text{width}(R(\nu_i)) \leq \text{width}(R(\nu))$. Thus $\text{width}(R(\nu_i))/\text{height}(R(\nu_i)) \leq \text{width}(R(\nu))/\text{height}(R(\nu)) \leq k$. □

**Case 2: $d(\nu_1) = d(\nu)$.** By construction of $\mathcal{T}^*$, $1/3 \cdot \text{weight}(\nu) \leq \text{weight}(\nu_1) \leq 2/3 \cdot \text{weight}(\nu)$. We now partition $R(\nu)$ into two subpolygons of the appropriate area with an axis-parallel cut orthogonal to the longest side of the axis-parallel bounding box of $R(\nu)$. The possible positions of this cut are limited by convexity, as specified in the following lemma.

**Lemma 5.** *Let $P$ be a convex polygon with $\text{width}(P) \geq \text{height}(P)$. We can partition $P$ with a vertical cut into two subpolygons $P_1, P_2$, where $\text{area}(P)/3 \leq \text{area}(P_i) \leq 2/3 \cdot \text{area}(P)$ (for $i \in \{1,2\}$), such that $\text{width}(P)/4 \leq \text{width}(P_i) \leq 3/4 \cdot \text{width}(P)$.*

PROOF. Assume, without loss of generality, that $\text{area}(P) = 1$. We cut the axis-parallel smallest enclosing rectangle of $P$ into three vertical slices l, c and r of relative widths $1/4$, $1/2$ and $1/4$, respectively. Let $P_x$ be the intersection of $P$ with slice $x \in \{l, c, r\}$. Without loss of generality, we assume $\text{area}(P_l) \geq \text{area}(P_r)$. The lemma then follows from $\text{area}(P_l) \leq 1/2$ and $\text{area}(P_r) \leq 1/3$, since we can always let the left subpolygon be the larger. It is not hard to see that the relative area of $P_l$ is maximal if $P$ is a triangle and the relative area of $P_r$ is maximal if $P$ is a rectangle, as shown. In the first case, $\text{area}(P_l) = 7/16 < 1/2$. In the latter, $\text{area}(P_r) = 1/4 < 1/3$. □

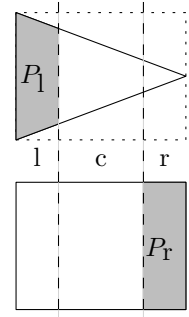

Clearly the number of non-axis-parallel edges of $R(\nu_1)$ and $R(\nu_2)$ is no more than the number of non-axis-parallel edges of $R(\nu)$. Since $d(\nu_i) \geq d(\nu)$, this implies $R(\nu_1)$ and $R(\nu_2)$ satisfy (Inv-1). As for (Inv-2), note that the cut does not introduce any new non-axis-parallel edges. It thus remains to prove the following lemma:

**Lemma 6.** *If $R(\nu_i)$ has two horizontal edges, then $\text{width}(R(\nu_i))/\text{height}(R(\nu_i)) \leq 4$ (for $i \in \{1,2\}$). Similarly, if $R(\nu_i)$ has two vertical edges, $\text{height}(R(\nu_i))/\text{width}(R(\nu_i)) \leq 4$.*

PROOF. We write $h$ for $\text{height}(R(\nu))$ and $w$ for $\text{width}(R(\nu))$ and define $h_i$ and $w_i$ similarly for $R(\nu_i)$. Assume, without loss of generality, that $h \leq w$, hence $R(\nu_i)$ is obtained from $R(\nu)$ by a vertical cut. Clearly, $h_i \leq h$. Moreover, by Lemma 5 we have $w_i \geq w/4$, hence $h_i/w_i \leq 4h/w \leq 4$.

If $R(\nu_i)$ has two horizontal edges, then so has $R(\nu)$, since no horizontal edges are introduced by the cut. Therefore $w/h \leq 4$. We know $h_i = h$ and obviously, $w_i \leq w$, hence $w_i/h_i \leq w/h \leq 4$. □

Lemma 3, together with the fact that $\max_{\nu \in \mathcal{T}^*} d(\nu) = \text{depth}(\mathcal{T})$ and (Inv-2), implies the result.

**Theorem 7.** *Every properly weighted tree of depth d can be represented by a convex treemap (polygonal partition) which has aspect ratio $O(d)$.*

### 4. Ortho-convex Treemaps

We describe a recursive algorithm for computing an orthoconvex treemap of constant aspect ratio for a properly weighted *binary* tree $\mathcal{T}$. If our original input tree is not binary, we can simply replace each node of degree $k > 2$ by a binary subtree with $k-1$ new internal nodes. Roughly speaking, at every step our algorithm partitions the tree under consideration into several "well-sized" pieces which can be drawn inside rectangles, L-, and S-shapes of constant aspect ratio. The main difficulty is that subtrees may be split over several such pieces. We have to make sure that the fragments corresponding to the same subtree end up bordering each other so that the region corresponding to the subtree is orthoconvex. We solve this problem by *marking* certain corners and nodes during the recursive process, and letting our drawing algorithm be guided by the marked corners and nodes.

Our algorithm uses *staircases*: polygons defined by a horizontal edge $uv$, a vertical edge $vw$, and an $xy$-monotone chain of axis-parallel edges connecting $u$ to $w$. The vertex $v$ is called the *anchor* of the staircase. At each recursive step we are given a rectangle $R$ with aspect ratio at most 8 and a tree $\mathcal{T}$. Exactly one node $\mu$ in $\mathcal{T}$ and exactly one corner of $R$ is *marked*. Initially $\mathcal{T}$ is the input tree, $R$ is the unit square, and the root of $\mathcal{T}$ and the bottom-right corner of $R$ are marked. We compute a treemap for $\mathcal{T}$ inside $R$ with the following properties:

(i) every leaf is drawn as a rectangle of aspect ratio at most 8, or as an L- or S-shape of aspect ratio at most 32;
(ii) every internal node is drawn as an orthoconvex polygon of aspect ratio at most 64;
(iii) the marked node $\mu$ as well as its ancestors are drawn as staircases whose anchors coincide with the marked corner of $R$.

The third property is not a goal in itself, but it is necessary to guarantee the other two. We now describe how to draw $\mathcal{T}$ inside $R$. Our algorithm distinguishes cases depending on the *relative weights* of certain nodes. The relative weight $\text{rel}(\nu)$ of a node $\nu$ is its weight as a fraction of the weight of the tree $\mathcal{T}$ currently under consideration. We partition the relative weights into four categories:

- *tiny nodes*: nodes $\nu$ such that $\text{rel}(\nu) < 1/8$;
- *small nodes* nodes $\nu$ such that $1/8 \leq \text{rel}(\nu) < 1/4$;
- *large nodes*: nodes $\nu$ such that $1/4 \leq \text{rel}(\nu) \leq 7/8$;
- *huge nodes*: nodes $\nu$ such that $\text{rel}(\nu) > 7/8$.

For a node $\nu$ we use $\mathcal{T}_\nu$ to denote the subtree rooted at $\nu$. Moreover, we write $p(\nu)$ for the parent of $\nu$ and $s(\nu)$ for the sibling of $\nu$.

**Lemma 8.** *Let $\nu$ be a non-tiny node in $\mathcal{T}$. Then $\mathcal{T}_\nu$ contains a non-tiny leaf or a node that is small or large.*

PROOF. All weights are relative to $\mathcal{T}$. If $\nu$ is small or large we are done, so assume $\nu$ is huge. Walk from $\nu$ down $\mathcal{T}_\nu$, always proceeding to the heavier child (breaking ties arbitrarily) until we reach a leaf or a non-huge node. Since the heavier child of a huge node is huge or large, this ends in a huge leaf or a large node. □

We now discuss the various cases that we distinguish, show how the algorithm handles them, and prove (using induction) that each case is handled correctly. In the base case $\mathcal{T}$ consists of a single leaf node, so its region is simply the rectangle $R$. This trivially satisfies conditions (i)–(iii). So now assume $\mathcal{T}$ has more than one node. In the following, whenever we mark a node in a tree that already has a marked node $\mu$, we implicitly assume that the mark is removed from $\mu$. In the description below—in particular in the figures illustrating our approach—we assume without loss of generality that the bottom-right corner of container $R$ is marked and that $\text{width}(R) \geq \text{height}(R)$.

**Case (a): $\mathcal{T}$ has a non-tiny marked node $\mu$.** By Lemma 8, $\mathcal{T}_\mu$ contains a node $\nu$ that is either a non-tiny leaf or a small or large internal node. Let $\mathcal{T}'$ be the tree obtained from $\mathcal{T}$ by removing $\mathcal{T}_\nu$ and contracting $s(\nu)$ into $p(\nu)$. In $\mathcal{T}'$ we mark $s(\nu)$ and in $\mathcal{T}_\nu$ we mark $\nu$. We distinguish two subcases.

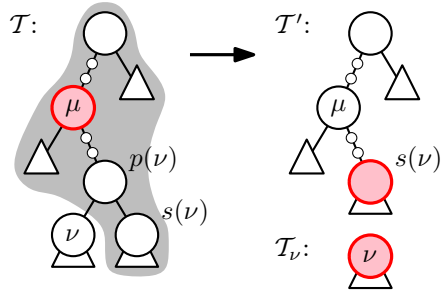

If $\text{rel}(\nu) \leq 7/8$ then we split $R$ into two subrectangles $R'$ and $R(\nu)$, one for $\mathcal{T}'$ and one for $\mathcal{T}_\nu$. We put $R(\nu)$ to the right of $R'$ and mark the bottom-right corner of both. Note that $R'$ and $R(\nu)$ have aspect ratio at most 8, according to Lemma 2. We recursively draw the trees in their respective rectangles.

If $\text{rel}(\nu) > 7/8$ then $\nu$ must be a leaf. We then draw $\nu$ as an L-shape $R(\nu)$ and recursively draw $\mathcal{T}'$ inside a rectangle $R'$ which is *similar* to $R$, whose top-left corner coincides with the top-left corner of $R$, and whose bottom-right corner is marked.

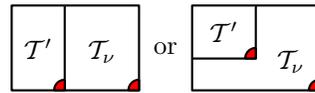

In both subcases properties (i)–(iii) hold. The recursive calls only draw leaf regions having property (i) and the L-shaped leaf in the second subcase has aspect ratio at most $64/7$.

Also, all internal nodes have property (ii). For the nodes that are not an ancestor of $\nu$ this follows by induction. For the ancestors of $\nu$, orthoconvexity follows from the fact that the recursive call on $\mathcal{T}'$ has property (iii), and the relative positions of $R(\nu)$ and the marked corner of $R'$. That is, every ancestor of $\nu$ is drawn as a staircase with its anchor 'glued' to the rectangle or L-shape $R(\nu)$. Since $\nu$ is non-tiny, we know $\text{area}(\hat{\nu}) \geq \text{area}(R)/8$, for all ancestors $\hat{\nu}$ of $\nu$. Moreover, since $\mathrm{asp}(R) \leq 8$, we have $\mathrm{area}(R) \geq \mathrm{area}(\sigma(R))/8$. Therefore, the regions for the ancestors of $\nu$ have aspect ratio at most 64.

The marked node $\mu$ and all of its ancestors are ancestors of the marked node $s(\nu)$ in $\mathcal{T}'$. Therefore, these are drawn as staircases in $R'$, by induction. The subtrees rooted at these nodes contain $\mathcal{T}_\nu$ in $\mathcal{T}$. A staircase whose anchor is 'glued' to the rectangle or L-shape $R(\nu)$ is a staircase again, with its anchor in the marked corner of $R$. Therefore, $\mu$ and all of its ancestors have property (iii).

**Case (b): $\mathcal{T}$ has a tiny marked node $\mu$ with an ancestor that is small or large.** Let $\mu^*$ be the lowest huge ancestor of $\mu$—since the root is huge, $\mu^*$ must exist—and let $\hat{\mu}$ be the child of $\mu^*$ on the path to $\mu$. Then $\hat{\mu}$ is small or large. We obtain $\mathcal{T}'$ by removing $\mathcal{T}_{\hat{\mu}}$ and contracting $s(\hat{\mu})$ into its parent. We mark $s(\hat{\mu})$ in $\mathcal{T}'$ and $\mu$ remains marked in $\mathcal{T}_{\hat{\mu}}$.

We split $R$ into two rectangles $R'$ and $R(\hat{\mu})$, one for $\mathcal{T}'$ and one for $\mathcal{T}_{\hat{\mu}}$. Since $\hat{\mu}$ is small or large, the aspect ratios of $R'$ and $R(\hat{\mu})$ are at most 8, according to Lemma 2. We put $R(\hat{\mu})$ to the right of $R'$, mark the bottom-right corner in both, and recursively draw $\mathcal{T}'$ in $R'$ and $\mathcal{T}_{\hat{\mu}}$ in $R(\hat{\mu})$.

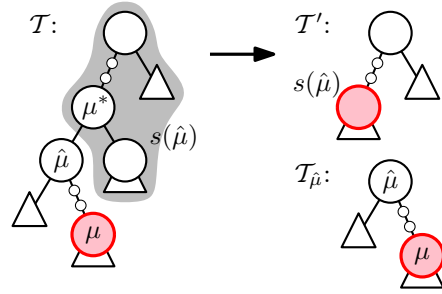

The leaf regions have property (i) by induction. Property (ii) holds as well. For the nodes that are not an ancestor of $\hat{\mu}$ this follows by induction. For the ancestors of $\hat{\mu}$, orthoconvexity follows from the fact that the recursive call on $\mathcal{T}'$ has property (iii), and the relative positions of $R'$ and the marked corner of $R(\hat{\mu})$. Because $\hat{\mu}$ is non-tiny, the regions for the ancestors of $\hat{\mu}$ have aspect ratio at least 64. Property (iii) follows from the fact that the recursive calls on $\mathcal{T}'$ and $\mathcal{T}_{\hat{\mu}}$ have property (iii) and from the relative position of $R'$ and the marked corners of $R(\hat{\mu})$ and $R'$.

**Case (c): $\mathcal{T}$ has a tiny marked node $\mu$ without small or large ancestors, but $\mathcal{T}$ has a large or huge leaf $\lambda$.** Define $\mu^*$ and $\hat{\mu}$ as in Case (b). Note that $\hat{\mu}$ must be tiny, since $\mu$ does not have small or large ancestors and $\mu^*$ is the lowest huge ancestor. Also note that $\lambda$ must be in the other subtree of $\mu^*$ (the one not containing $\hat{\mu}$ and $\mu$). Now we get $\mathcal{T}'$ from $\mathcal{T}$ by removing $\mathcal{T}_{\hat{\mu}}$ and $\lambda$, and

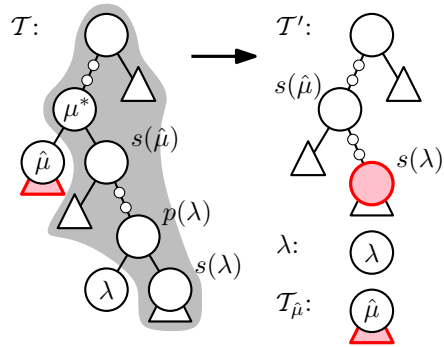

contracting $s(\hat{\mu})$ and $s(\lambda)$ into their parents. We mark $s(\lambda)$ in $\mathcal{T}'$ and keep $\mu$ marked in $\mathcal{T}_{\hat{\mu}}$. We draw $\mathcal{T}_{\hat{\mu}}$ in a rectangle $R(\hat{\mu})$ similar to $R$, in the bottom-right of $R$, with its bottom-right corner marked.

If $\mathcal{T}'$ is tiny we draw it as a rectangle $R'$ similar to $R$, in the top-left of $R$; otherwise $\mathcal{T}'$ is drawn as a rectangle $R'$ on the left side of $R$. Note that 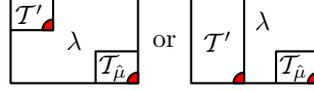
in the latter case the aspect ratio of $R'$ is at most 8. In both cases we mark the bottom-right corner of $R'$. The region $R(\lambda)$ for $\lambda$ is drawn in between $R'$ and $R(\hat{\mu})$, which means it is either an S-shape or an L-shape.

All leaves in $\mathcal{T}'$ and $\mathcal{T}_{\hat{\mu}}$ have property (i) by the induction hypothesis. Since $\lambda$ is large or huge, the aspect ratio of $R(\lambda)$ is at most 32. Properties (ii) and (iii) follow using the induction hypothesis on $\mathcal{T}'$ and $\mathcal{T}_{\hat{\mu}}$, similarly as before.

**Case (d): $\mathcal{T}$ has a tiny marked node $\mu$ without small or large ancestors, and $\mathcal{T}$ has no large or a huge leaf.** Define $\mu^*$ and $\hat{\mu}$ as in Case (b) and (c). As in Case (c), $\hat{\mu}$ is tiny, so $\mathcal{T} \setminus \mathcal{T}_{\hat{\mu}}$ has relative weight at least $7/8$. We search for a node $\hat{\nu}$ such that $\text{rel}(\hat{\nu}) \leq 6/8$. This can be done by descending from $\mu^*$, always proceeding to the heavier child, until we reach such a node. Observe that $\text{rel}(\hat{\nu}) \geq 3/8$. Let $\mathcal{T}'$ be obtained by removing $\mathcal{T}_{\hat{\mu}}$ and $\mathcal{T}_{\hat{\nu}}$, marking $s(\hat{\nu})$ and contracting it in its parent. By construction, $\mathcal{T}'$ is not tiny. Let $\nu$ in $\mathcal{T}_{\hat{\nu}}$ be a small node; such a node $\nu$ exists since there are no large or huge leaves and the relative weight of $\hat{\nu}$ is at least $3/8$. Let $\mathcal{T}'_{\hat{\nu}}$ be the tree obtained from $\mathcal{T}_{\hat{\nu}}$ by removing $\mathcal{T}_{\nu}$, marking $s(\nu)$ and contracting it into its parent. Let $\mathcal{T}_r$ be constructed by joining $\mathcal{T}_{\hat{\mu}}$ and $\mathcal{T}_{\nu}$ with a new root $r$, as shown. 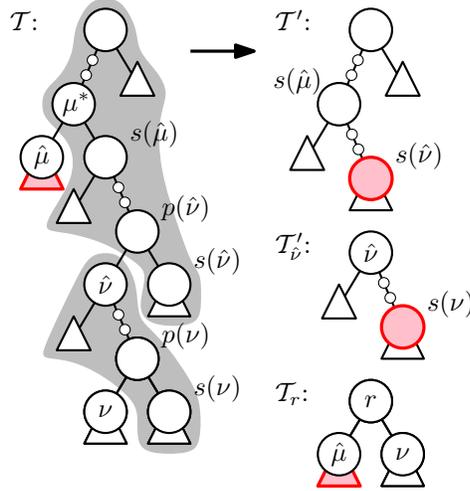

We partition $R$ into three rectangles, one for $\mathcal{T}'$, one for $\mathcal{T}'_{\hat{\nu}}$, and one for $\mathcal{T}_r$. We mark them as shown in the figure below, and recurse. Since all three subtrees are non-tiny the rectangles on which we recurse have aspect ratio at most 8.

All leaf regions have property (i) by induction. Moreover, all internal nodes have property (ii). For the nodes that are not an ancestor of $\nu$ this follows by induction. For nodes on the path from $\hat{\nu}$ to $\nu$ we can argue as follows. $R(\hat{\mu})$ is a staircase 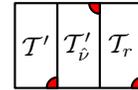
anchored at the bottom-right corner of the rectangle of $\mathcal{T}_r$, and so $R(\nu)$ is a staircase anchored at the opposite corner. Because the top-right corner of the rectangle of $\mathcal{T}'_{\hat{\nu}}$ is marked, this implies that the nodes from $\hat{\nu}$ to $\nu$ are drawn as orthoconvex polygons. A similar argument applies to the ancestors of $\hat{\nu}$. The nodes that are not an ancestor of $\nu$ have aspect ratio at most 64 by induction. The ancestors of $\nu$ are not tiny, since $\nu$ is not tiny, and $\text{asp}(R) \leq 8$, hence all the regions have aspect ratio at most 64.

It remains to argue that $\mu$ as well as all of its ancestors have property (iii).

For $\mu$ and its ancestors in $\mathcal{T}_{\hat{\mu}}$ this follows by induction. For $\mu^*$ and its ancestors this follows from the fact that the regions of each of these nodes contain the rectangle of $\mathcal{T}'_{\hat{\nu}}$ and $\mathcal{T}_r$, and the induction hypothesis on $\mathcal{T}'$.

**Theorem 9.** *Every properly weighted tree can be represented by an orthoconvex treemap in which all leaves are drawn as rectangles, L-shapes, or S-shapes, all internal nodes are drawn as orthoconvex polygons, and the aspect ratio of any of these regions is $O(1)$.*

## 5. Single-level Treemaps

Here we consider the special case that $\text{depth}(\mathcal{T}) = 1$, that is, single-level treemaps. We first show that if we restrict the regions to be rectangles, minimizing the aspect ratio is NP-hard. Moreover, in that case even the optimal drawing can have unbounded aspect ratio: consider a single-level tree with two leafs and let the weight of one of the leafs tend to zero. Then, its region is a very skinny rectangle.

Therefore, we describe a drawing procedure in which we also allow regions to be drawn as L-shapes. We show that the aspect ratio bound we obtain is nearly tight and much better than for the general case.

### 5.1. Tree mapping is NP-hard

In this section we prove that *tree mapping* is strongly NP-hard. Tree mapping is the problem of finding, for a given properly weighted single-level tree $\mathcal{T}$, a treemap such that all regions $R_1, R_2, \ldots, R_n$ are rectangles and $\max(1 \leq i \leq n : \text{asp}(R_i))$ is minimized. Our reduction is from *square packing* which was proven to be strongly NP-complete by Leung *et al.* [12]. Square packing is the problem of deciding for a given packing square $S$ and a set of squares $\mathcal{S}$, each having integer side lengths, if there is an orthogonal (i.e. axis-aligned) packing of $\mathcal{S}$ in $S$. Assume that we are given a square packing instance SP: $(S, \mathcal{S} = \{s_1, s_2, \ldots, s_l\})$. We create a tree mapping instance TM: $\mathcal{T} = \{w_1, w_2, \ldots, w_n\}$ with

$$n = l + \text{area}(S) - \sum_{i=1}^{l} \text{area}(s_i) \quad \text{and} \quad w_i = \begin{cases} \text{area}(s_i)/\text{area}(S) & \text{if } 1 \leq i \leq l, \\ 1/\text{area}(S) & \text{if } l < i \leq n. \end{cases}$$

**Lemma 10.** *SP has a solution if and only if TM has aspect ratio 1.*

PROOF. Assume that TM has aspect ratio 1. Then all weights are drawn as squares. Leaving out the lowest $n - l$ weights immediately gives a solution for SP. Now, assume that SP has a solution. Since all squares have integer side lengths, we can align all squares with the unit grid. The lowest $n - l$ weights of TM occupy exactly one grid cell each. Therefore, these can be drawn as squares as well and TM has aspect ratio 1. □

**Theorem 11.** *Tree mapping is strongly NP-hard.*

PROOF. By Lemma 10 we have a solution to a square packing instance iff we have a tree mapping solution of aspect ratio 1. Since square packing is strongly NP complete, we may assume that the size of $S$ is polynomial in $l$. Then, $n$ is polynomial in $l$ as well. Therefore, the reduction can be done in polynomial time. □

*5.2. Allowing L-shapes*

We show that for single-level orthoconvex treemaps we can obtain even nicer shapes and better aspect ratio than in the general case. Specifically, in the following we describe a recursive drawing procedure that uses L-shapes in addition to rectangles. The resulting treemap has aspect ratio at most $2+2\sqrt{3}/3 \approx 3.15$. Unlike for hierarchies, we cannot convert our input into a binary tree as a preprocessing step, since the desired structure of a subtree rooted at node $\nu$ depends on the aspect ratio of $R(\nu)$. Hence, we construct a binary tree $\mathcal{T}'$ on-the-fly. We draw a treemap for $\mathcal{T}'$ inside the unit square with the following properties:

(i) every leaf is drawn as a rectangle or an L-shape and has aspect ratio at most $2 + 2\sqrt{3}/3 \approx 3.15$;
(ii) every internal node is drawn as a rectangle and has aspect ratio at most $1 + \sqrt{3} \approx 2.73$.

Property (ii) is not a goal by itself, but is necessary to guarantee the first. In a generic step, we are given a node $\nu$ with associated region $R(\nu)$. If $\nu$ is a leaf, we are done, so assume that $\nu$ has $k \geq 2$ children $\nu_1, \nu_2, \ldots, \nu_k$. Then, $R(\nu)$ is a rectangle by (ii). Assume that $\nu_1$ is the heaviest child. We distinguish 3 cases:

**Case 1:** $\mathrm{rel}(\nu_1) < (3-\sqrt{3})/4$. Since $(3-\sqrt{3})/4 < 1/3$, we can partition children$(\nu)$ into subsets $H_1, H_2$, such that weight$(H_2) \leq$ weight$(H_1) \leq (3-\sqrt{3}) \cdot$ weight$(\nu)/2$ (Lemma 1). We turn $\nu$ 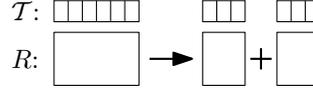 into a binary node with children $\mu_1, \mu_2$, with children from $H_1$ and $H_2$, respectively. We partition $R(\nu)$ into two rectangles $R(\mu_1)$ and $R(\mu_2)$ by cutting the longer side of $R(\nu)$, according to the weights of $\mu_1$ and $\mu_2$. We know weight$(H_2) \geq (1-(3-\sqrt{3})/2) \cdot$ weight$(\nu)$. With Lemma 2 it follows that both $R(\mu_1)$ and $R(\mu_2)$ meet (i) or (ii).

**Case 2:** $(3-\sqrt{3})/4 \leq \mathrm{rel}(\nu_1) < \mathrm{asp}(R(\nu)) \cdot (3-\sqrt{3})/4$. We turn $\nu$ into a binary tree with children $\nu_1$ and a new node $\mu$, having children children$(\nu)\setminus$ 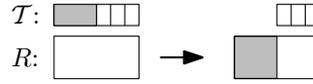 $\nu_1$. We partition $R(\nu)$ into two rectangles $R(\nu_1)$ and $R(\mu)$ by cutting the longer side of $R(\nu)$, according to the weights of $\nu_1$ and $\mu$. By Lemma 2 $R(\nu_1)$ meets (i) and $R(\mu)$ meets (i) or (ii).

**Case 3:** $\mathrm{rel}(\nu_1) \geq \mathrm{asp}(R(\nu)) \cdot (3-\sqrt{3})/4$. We turn $\nu$ into a binary tree with children $\nu_1$ and a new node $\mu$, having children children$(\nu) \setminus \nu_1$. 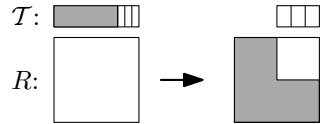 We partition $R(\nu)$ into a rectangle $R(\mu)$, similar

to $R(\nu)$ and with area proportional to weight($\mu$),
and an L-shape $R(\nu_1)$, with area proportional to weight($\nu_1$). Since $R(\mu)$ is similar with $R(\nu)$, it trivially meets (i) or (ii). By definition, the aspect ratio of $R(\nu_1)$ is at most $\mathrm{asp}(R(\nu))/\mathrm{rel}(\nu_1)$. A standard calculation shows that $R(\nu_1)$ meets (i).

Property (i) now implies the following theorem.

**Theorem 12.** *Every properly weighted single-level tree $\mathcal{T}$ can be represented by an orthoconvex treemap which uses only rectangles and L-shapes and has aspect ratio at most $2 + 2\sqrt{3}/3 \approx 3.15$.*

This result is nearly tight, a lower bound of $\approx 3.13$ is given in the following section.

*5.3. Lower bound*

Consider a single-level tree $\mathcal{T} = \{w_1, w_2, w_3, w_4\}$, where $w_1 = w_2 = w_3 = x$ for some $x < 1/3$. As a result, $w_4 = 1 - 3x$. We are particularly interested in the cases where $x \approx 1/3$, that is, three regions are large and one is small. Then, we can make a few observations:

- If a region corresponding to any of the high weights touches two opposite sides of the container, its aspect ratio is $1/x$, since the container is the smallest enclosing square.
- We only allow L-shapes whose remainder is a rectangle, as shown in the top figure. The bottom two figures show L-shapes that are disallowed. Therefore, an allowed L-shape always touches all sides of the container $C$ it is in. Consider such a subcontainer $C$ and let $\mathcal{T}'$ be the set of weights that are not drawn in $C$. Then, $|\mathcal{T}'| \leq 2$, since at least two weights are drawn in $C$. The entire container has four corners, so at least $C$ or one of the weights in $\mathcal{T}'$ touches two opposite sides of the container, by the pigeon hole principle. 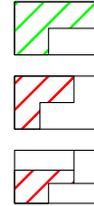
- As a result of the previous two items, if we use an L-shape, at least one of the regions touches two opposite sides of the container and the aspect ratio is at least $1/x$. Hence, the only way of getting an aspect ratio better than $1/x$ is by not using L-shapes and with no rectangle touching two opposite sides of $C$. This drawing has aspect ratio $(1-2x)^2/(1-3x)$. 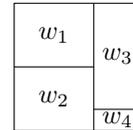

In order to find a worst-case optimal solution, we want to maximize the minimum of $1/x$ and $(1-2x)^2/(1-3x)$. That is, we solve the equation $1/x = (1-2x)^2/(1-3x)$, which gives aspect ratio $6/(2 + \sqrt[3]{3\sqrt{57} - 1} - \sqrt[3]{3\sqrt{57} + 1}) \approx 3.13$.

## 6. Conclusions and Open Problems

Treemaps are a popular visualization tool for hierarchical data and aspect ratio is an important quality criterion for them. We have settled two main theoretical questions concerning treemaps: First, we presented an algorithm for computing convex treemaps whose aspect ratio is $O(d)$, where $d$ is the depth of the input hierarchy; this is asymptotically optimal. Second, we showed that any input 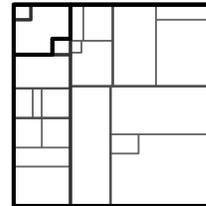
hierarchy admits an orthoconvex treemap of constant aspect ratio. We also implemented the algorithms. Preliminary experiments show that the aspect ratios remain significantly below their provable worst-case bounds, and that orthoconvex treemaps hardly ever use S-shapes. (The figure on the right was carefully constructed to illustrate S-shapes.)